\documentclass[notitlepage, twoside,english,nofootinbib,preprintnumbers,aps,11pt,showpacs]{revtex4-1} \usepackage[a4paper, margin=1.5in]{geometry} \usepackage{amssymb} \usepackage{commath}
\usepackage{amsmath}
\usepackage{url}
\usepackage{float}
 \usepackage[utf8]{inputenc} \usepackage[T1]{fontenc} \usepackage{lmodern}
\usepackage{braket}
\usepackage{graphicx}
\usepackage{pictexwd,dcpic}
\usepackage{braket}
\usepackage{atbegshi,picture, afterpage}
\usepackage{lipsum}
\newcommand{\be}{\begin{equation}}
\newcommand{\ee}{\end{equation}}
\newcommand{\ba}{\begin{align}}
\newcommand{\ea}{\end{align}}
\begin{document}
\title{Dressed metric predictions revisited}
\author{Wojciech Kamiński}
\email[]{Wojciech.Kaminski@fuw.edu.pl}
\author{Maciej Kolanowski}
\email[]{mp.kolanowski@student.uw.edu.pl}
\author{Jerzy Lewandowski}
\email[]{Jerzy.Lewandowski@fuw.edu.pl}
\affiliation{\vspace{6pt} Institute of Theoretical Physics, Faculty of
  Physics, University of Warsaw, Pasteura 5, 02-093 Warsaw, Poland}
  \date{\today}
\begin{abstract}
    It was recently shown that the volume operator of loop quantum cosmology (LQC) and all its positive powers are ill-defined on physical states. In this paper, we investigate how it effects predictions of cosmic microwave background (CMB) power spectra obtained within dressed metric approach for which expectations values of $\hat{a}$ are the key element. We find that almost every step in the procedure is ill-defined and relies heavily upon a (seemingly premature) numerical truncation. Thus, it suggests that more care is needed in making predictions regarding pre-inflationary physics. We propose a new scheme which contains only well-defined quantities. The surprising agreement of the hitherto models with observational data, especially at low angular momenta $l$ is explained.
\end{abstract}
\maketitle
\section{Introduction}
It was recently shown \cite{Kaminski:2019tqo} that neither the scale factor  operator's $\hat{a}$ nor its positive powers $\hat{a}^\epsilon$ domains are preserved under the time evolution in the most important models of loop quantum cosmology (LQC)  including the Ashtekar-Singh-Pawłowski (ASP) model, as well as the solvable LQC (sLQC).  As a consequence, the expectation values like $\langle \hat{a}(T) \rangle$ are ill-defined for a generic semiclassical state at a generic instant of time.  On the other hand, the LQC models are used to predict the power spectra of cosmological microwave background (CMB) and related quantities (see e.g. \cite{Bonga:2015kaa, Bonga:2015xna, Agullo:2015tca}) and their results are in excellent agreement with empirical data \cite{Akrami:2018odb}.
The natural questions arises, whether  those predictions can be really independent of the aforementioned issue? Is there any subtle mechanism that makes the infinities coming from  $\langle \hat{a}(T) \rangle$  cancel? Or is  a better understanding of the numerical results necessary? In this paper we look from this angle at the dressed metric approach which is a perturbation scheme for the pre-inflationary cosmology.
\section{Dressed metric approach}
The perturbation scheme for LQC coupled to inhomogeneous scalar field perturbations  was initiated in \cite{Ashtekar:2009mb} and developed in \cite{Agullo:2013ai, Agullo:2012sh} for applications to CMB power spectra and other fields. 

\subsection{The quantum cosmological spacetime}
The homogeneous isotropic  background  quantum geometry is given by the LQC  models, and takes the form of operator valued quantum spacetime metric tensor
\begin{equation}
\widehat{ds^2}\ =\ -\hat{H}_o^{-1} \hat{a}^6(T)\hat{H}_o^{-1}dT^2  + \hat{a}^2(T)dx^idx^i . 
\end{equation}
where 
\begin{equation}
\hat{a}(T) ={\rm exp}\left(\frac{i}{\hbar}\hat{H}_oT\right)\hat{a}\, {\rm exp}\left(-\frac{i}{\hbar}\hat{H}_oT\right).
\end{equation}
The effective quantum Hamiltonian operator $\hat{H}_o$ and the scale operator $\hat{a}$ are defined in a suitable Hilbert space. The background matter consists in spatially  homogeneous scalar field $T$ of the quantum  momentum  
\begin{equation}
\hat{\Pi}\ =\ \pm \hat{H}_o.  
\end{equation}
The recent result \cite{Kaminski:2019tqo} that motivated the current letter is that for the main LQC models, for every generic state  of the quantum cosmological spacetime, that is a state $\Psi$ such that
 \begin{equation}
\hat{H}_o\Psi \not=0
\end{equation} 
the expectation value of the scale operator $\hat{a}(T)$ suffers pathological property, namely  
 \begin{equation}\label{infty}
\langle \hat{a}(T)\rangle < \infty \ \ \ \ \Rightarrow\ \ \ \ \langle \hat{a}^\epsilon(T+\tau)\rangle = \infty,
\end{equation} 
for every finite interval $\tau$ and arbitrary real power $\epsilon>0$,  where $\langle \cdot \rangle$ are expectations values evaluated at any background state $\Psi$.      

The first and obvious consequence  is that the expected (formally classical)  metric tensor  $\langle \widehat{ds^2}\rangle$ is not well defined.  What about the effective classical  metric tensor  (the dressed metric) 
\begin{equation}
    d\Tilde{s}^2 = - \Tilde{N}^2 dT^2 + \Tilde{a}^2 dx^i dx^i, \label{dressed_metric}
\end{equation}
that is sensed by quantum test fields?  It is somewhat different, defined as   
\begin{align}
    \Tilde{a}^4 &= \frac{\langle \hat{H}_o^{-\frac{1}{2}} \hat{a}^4 (T) \hat{H}_o^{-\frac{1}{2}}  \rangle}{\langle \hat{H}_o^{-1} \rangle} \\
    \Tilde{N} &= \hbar \langle \hat{H}_o^{-\frac{1}{2}} \hat{a}^4 (T) \hat{H}_o^{-\frac{1}{2}}  \rangle^{\frac{3}{4}} {\langle \hat{H}_o^{-1} \rangle}^{\frac{1}{4}} 
\end{align}
however, the problem persits.  
\subsection{Equations of motion}
Let us assume for a moment that dressed metric is well-defined object, it means all expectation values are finite. Then, one can consider  either scalar $\mathcal{Q}$ or tensorial $\mathcal{T}$ perturbations which are evolved using this dressed metric. Following \cite{Agullo:2016tjh} $\mathcal{Q}$ represents Mukhanov variable. To be more precise, $\mathcal{T}$ satisfies wave equation with respect to the metric \eqref{dressed_metric}:
\begin{equation}\label{nabla}
    \Tilde{\nabla}^\mu \Tilde{\nabla}_\mu \mathcal{T} = 0
\end{equation}
whereas equation for $ \mathcal{Q}$ contains the additional potential term:
\begin{equation}
   \Tilde{U} = \frac{\langle \hat{H}_o^{-\frac{1}{2}} \hat{a}^2 (T) \hat{U} \hat{a}^2(T)\hat{H}_o^{-\frac{1}{2}}  \rangle}{\langle \hat{H}_o^{-\frac{1}{2}} \hat{a}^4(T) \hat{H}_o^{-\frac{1}{2}}  \rangle}.
\end{equation}
Since we are working on a homogeneous background, it is convenient to take the spatial Fourier transform, we will denote is component by $\mathcal{T}_k$ and $\mathcal{Q}_k$. We can introduce conformal time $\eta$ as
\begin{equation}\label{conf}
    d \eta = \frac{\Tilde{N}}{\Tilde{a}} dT = \hbar \langle \hat{H}_o^{-\frac{1}{2}} \hat{a}^4 (T) \hat{H}_o^{-\frac{1}{2}}  \rangle^{\frac{1}{2}} \langle \hat{H}_o \rangle^\frac{1}{2} dT = \hbar \Tilde{a}^{2} \langle \hat{H}_o^{-1} \rangle dT
\end{equation}
in which evolution equation is particularly simple:
\begin{align} 
    \mathcal{T}''_k + 2 \frac{\Tilde{a}'}{\Tilde{a}} \mathcal{T}'_k + k^2 \mathcal{T}_k &= 0, \label{conf_eomT} \\
    \mathcal{Q}''_k + 2 \frac{\Tilde{a}'}{\Tilde{a}} \mathcal{Q}'_k + (k^2 + \Tilde{U}) \mathcal{Q}_k &= 0 \label{conf_eomQ},
\end{align}
  We turn now to the case  when (\ref{infty})  holds. It is a simple generalization of \cite{Kaminski:2019tqo} to any positive power of volume (and thus also $\hat{a}$) operator. It is true for $k=0, \Lambda = 0$ standard LQC with massless scalar field (at least for APS \cite{Ashtekar:2006wn}, MMO \cite{MartinBenito:2009aj}, sLQC \cite{Ashtekar:2007em} and sMMO \cite{MenaMarugan:2011me}).
Obviously,  the equations (\ref{nabla}) and  (\ref{conf}) written above  do not make sense any more (at least not for more than one instance of time $T$). However, one could hope that the final equations $($\ref{conf_eomT}-\ref{conf_eomQ}$)$ may  somewhat magically, due to some cancelling (at least conceivable) consist of well-defined terms.  
In order to check it out, we have to rewrite the  equations of motion (\ref{conf_eomT}-\ref{conf_eomQ}) in terms of the scalar field  time $T$ used in our model. 
We have $\partial_\eta = \frac{1}{\hbar\langle \hat{H}_o^{-1} \rangle} \frac{1}{\Tilde{a}^2} \partial_T$. By a simple application of a chain rule, we obtain wave equation in the time $T$:
\begin{equation}\label{T}
    \mathcal{T}_{k,TT} + \langle \hat{H}_o^{-1} \rangle^2 \Tilde{a}^4 k^2 \mathcal{T}_k = 0.
\end{equation}
The scalar perturbation $\mathcal{Q}_k$ satisfies the similar equation 
\begin{equation}\label{Q}
    \mathcal{Q}_{k,TT} + \langle \hat{H}_o^{-1} \rangle^2 \Tilde{a}^4 (k^2 + \tilde{U})\mathcal{Q}_k = 0.
\end{equation}
Obviously,  the term $\Tilde{a}^4$ cannot be regularized in any natural manner.

\subsection{The choice of vacuum}
Until now, we were treating $\mathcal{T}$ and $\mathcal{Q}$ as a classical variables. In fact they are Fock quantized. Suitably selected  solutions  $\mathcal{T}_k$ and $\mathcal{Q}_k$  of (\ref{T}, and \ref{Q}) serve as  coefficients to the annihilation and creation operators $\hat{A}_{\vec{k}}$ and $\hat{A}_{\vec{k}}^\dagger$, respectively:
\begin{equation}
    \hat{\mathcal{T}} (T, \vec{x}) = \int \frac{d^3 k}{(2\pi)^3} \left( \hat{A}_{\vec{k}} \mathcal{T}_k + \hat{A}^\dagger_{-\vec{k}} \mathcal{T}_k^\star
    \right) e^{i \vec{k} \cdot \vec{x}}
\end{equation}
Obviously, different choices of $\mathcal{T}_k$ lead to different $\hat{A}_{\vec{k}}$ and thus different (possibly not unitarly equivalent) vacua. Our notation in denoting $\mathcal{T}_{\vec{k}}$ as $\mathcal{T}_k$ already suggests that the chosen vacuum state is adapted to the geometry and thus invariant under euclidean group. This however is not enough, one needs also to define an asymptotic behaviour of $\mathcal{T}_k$. In the usual, slow-roll inflation one can introduce the notion of Bunch-Davies state. Unfortunately, it is not the case near the Bounce (on which in particular Hubble constant vanishes). It was argued in \cite{Agullo:2016tjh} that it is physically reasonable to assume that the state satisfies adiabatic condition at the $4\textrm{th}$ order. In particular, it is the minimal requirement to assure the existence of the stress--energy tensor for perturbations. This fixes $\mathcal{T}_k$ up to the correction of order $\left(\frac{\Tilde{a}}{k} \right)^{\frac{9}{2}}$. Since $\Tilde{a} = \infty$, also this requirement is not well-posed. 
\section{Alternative approach to effective spacetime}
In this section we discuss more general attempt to describe an evolution of a perturbation on a quantum FLRW background. The main difference lies in the fact that in this method lapse function $N$ is not chosen to be $\frac{a^3}{p_\phi}$ and we obtain the whole family of Born-Oppenheimer-like dynamics for different $N$. By a (highly non-canonical) appropriate choice of lapse, we can obtain equations of motion for perturbation which contain only well-defined quantities. We assume, in the usual spirit of dressed metric dynamics, that several terms are negligible. Self-consistency of that assumption with numerical calculations can give further insight into applicability of a given approximation. \newline
For simplicity we work only with a free scalar field. Introduction of a potential and tensor modes is straightforward.
\subsection{Explicit cutoff approach}
In this section we will consider an alternative to the usual evolution of perturbations on quantum background. It leads to mathematically well-defined dressed metric. As a price, we need to introduce some ambiguities along the way. However, they were in fact present in the effective dynamics from the very beginning, our derivation merely makes them explicit. Moreover, it seems that for a large class of possible choices, obtained evolution equations are (almost) unambiguous. \newline
For simplicity, we are interested only in free, scalar field. Inclusion of the potential or tensor modes is straightforward. \\
We use the following ansatz for the state in the interaction picture:
\begin{equation}
    \Psi_{int}(v, Q_k, T) = \Psi_o (v, T_0) \otimes \Psi_p (Q_k, T), \label{ev_int}
\end{equation}
where $\Psi_o$ is a quantum geometry solving hamiltonian constraint. Separability of state corresponds to the assumption that backreaction can be ignored. One is led to the following Schrodinger-like equation (see (4.7) in \cite{Ashtekar:2009mb}):
\begin{equation}
    l^{-3} \Psi_o \otimes i \hbar \partial_T \Psi_p = \frac{1}{2} \left[  \hat{H}_o^{-1} \Psi_o \right] \otimes \left[ \hat{p}_k^2 \Psi_p \right] + \frac{1}{2} \left[ k^2 \hat{H}_o^{-\frac{1}{2}} \hat{a}^4 \hat{H}_o^{-\frac{1}{2}} \Psi_o \right] \otimes \left[ \hat{Q}_k^2 \Psi_p \right]
    \label{ev}
\end{equation}
If we were following previous attempts precisely, the next step would be to take scalar product in geometrical sector of our theory with $\Psi_0$. Unfortunately, as we have seen, obtained expressions are ill-defined. Instead, we can take scalar product with a vector $\hat{A} \Psi_o$. If Eq. \eqref{ev} held exactly, it would give us the same dynamics. Obviously, it is not the case and so we obtain a whole family of possible equations: 
\begin{equation}
    l^{-3} \langle \hat{A} \rangle i \hbar \partial_T \Psi_p = \frac{1}{2} \langle \hat{A}  \hat{H}_o^{-1} \rangle \hat{p}_k^2 \Psi_p  + \frac{1}{2}  k^2 \langle \hat{A} \hat{H}_o^{-\frac{1}{2}} \hat{a}^4 \hat{H}_o^{-\frac{1}{2}} \rangle \hat{Q}_k^2 \Psi_p.  \label{ev_A}
\end{equation}
Two problems rise immediately. First of all, there is an obvious ambiguity in the choice of an $\hat{A}$. Second of all, since operators in brackets are in general not even symmetric, their expectations values are going to be complex and so evolution of $\Psi_{pert}$ is not unitary. We will see that for semiclassical background, the latter issue does not matter. One could also hope it is true also for the former. \newline
Let us assume that $\Psi_o$ for some late time  $T$ (e.g. in the current universe) satisfies the following semiclassical property:
\begin{equation}
    \langle \hat{A} \hat{B} \rangle \approx \langle \hat{A} \rangle \langle \hat{B} \rangle \label{semic}
\end{equation}
for such $\hat{A}$, $\hat{B}$ that both sides are well-defined. It is well-known that LQC evolution preserves this condition even up to the Bounce. Let us take in Eq. \eqref{ev_A} $\hat{A} = f(\hat{a}) H_o^{\frac{1}{2}}$, where $\hat{f}$ is some function regular at $0$ and vanishing fast enough at infinity. Now we can approximate:
\begin{align}
    \begin{split}
        \langle \hat{A} \rangle &\approx \langle f(\hat{a}) \rangle \langle \hat{H}_o \rangle^{\frac{1}{2}} \\
        \langle \hat{A} H_0^{-1} \rangle &\approx \langle f(\hat{a}) \rangle \langle \hat{H}_o \rangle^{-\frac{1}{2}} \\
        \langle \hat{A} H_o^{-\frac{1}{2}} \hat{a}^4 \hat{H}_o^{-\frac{1}{2}} \rangle &\approx \langle f(\hat{a}) \hat{a}^4 \rangle \langle H_o \rangle^{-\frac{1}{2}}.
    \end{split}
\end{align}
We have assumed that $H_o^{\frac{1}{2}} H_o^{-\frac{1}{2}} = 1$. In general, it is not true due to the possible problems with domain of those operators. One should understand thus that by e.g. $\hat{A}  \hat{H}_o^{-\frac{1}{2}} \hat{a}^4 \hat{H}_o^{-\frac{1}{2}}$ we in fact understand simply $f(\hat{a}) \hat{a}^4  \hat{H}_o^{-\frac{1}{2}}$. This equality holds for matrix elements on a dense subspace of states  so we define $\hat{A} \hat{H}_o^{\frac{-1}{2}} \hat{a}^4 \hat{H}_o^{-\frac{1}{2}}$ space as a closure. \\
Any imaginary part vanishes in the semiclassical limit and thus unitarity is restored. Our choice of $\hat{A}$ corresponds to the free massless scalar field living on the following background:
\begin{equation}
    d\tilde{s}^2 = - \tilde{N}^2 dT^2 + \tilde{a}^2 dx^i dx^i, 
\end{equation}
where:
\begin{align}
    \begin{split}
        \tilde{a} &= \left( \frac{\langle f(\hat{a})\hat{a}^4 \rangle}{ \langle f(\hat{a} \rangle} \right)^{\frac{1}{4}} \\
        \tilde{N} &= l^3 \langle H_o^{-1} \rangle \langle f(\hat{a}) \rangle \tilde{a}^3. \label{dressed_corrected}
    \end{split}
\end{align}
In particular, it seems that $f(a) = a_0^4 \left(a^2 + a_0^2 \right)^{-2}$ with an appropriate choice of $a_0$ would work. From the final expressions above, it should be clear that $f(a)$ can be interpreted as a smooth cutoff in the volume. It has an advantage in comparison with a sharp cutoff used before such that it is state independent and it should work for all ranges of the volume. One could also hope that \eqref{semic} is satisfied.
\subsection{Comparison with previous works}
It would be of interest to compare this with results obtained previously in the literature. Unfortunately, our result relies heavily upon the assumption that the universe is semiclassical. It is in some sense in the conflict with the original idea that one could use effective evolution of a test field to probe higher moments of $\hat{a}$ operator. Here, we explicitly neglect at least some correlations. It would be of interest to check previous numerical calculations for signs of breaking from semiclassical regime, even so we expect smooth cutoff to be better.\\ 
So far, in numerical studies no divergences were detected which suggests that they are extremely small in agreement with analytical predictions \cite{Kaminski:2019tqo}. It is possible that just discovered infinities become visible way beyond first quantum corrections and thus one could still try to probe them.
\section{Summary} Recently, a crucial problem in the dynamical evolution of the expectation values of the quantum scale factor operator in some models of LQC was discovered  \cite{Kaminski:2019tqo}, namely the property (\ref{infty}). In the current paper  physical consequences of that result were investigated, in particular, whether  those  LQC models can be used to predict the cosmic microwave background (CMB) power spectra. Obviously, given (\ref{infty}), the dynamics  does not allow to define expectation values of quantum spacetime metric coefficients.  Moreover, an effective, semiclassical  dressed spacetime metric sensed by quantum fields propagating in a semiclassical state of LQC can not be defined either. Finally and most importantly, quantum test fields can not be defined either.  There is no chance for any canceling of infinities in the dressed scale factor used in the corresponding equations.  
Therefore, the  conclusion is, that  predictions on the cosmic microwave background (CMB) power spectra for the LQC models that have the property (\ref{infty}) is not possible. However, we proposed a generalized dressed metric approach \eqref{dressed_corrected} which is (at least in its final expressions) free of those mathematical issues. We conjecture that this could be seen also as a transformation from a theory with one lapse to another. \\
A more optimistic conclusion for Loop Quantum Cosmology is, that the  Big Bounce property is insensitive of the problem (\ref{infty})  because the energy density operator 
\begin{equation}
\hat{\rho}(T) = \frac{1}{2} \hat{a}^{-2}(T)\hat{H}_o^2 \hat{a}^{-2}(T). 
\end{equation} 
is  bounded.              
\begin{acknowledgments}
We thank Parampreet Singh. JK and WK thank for support Polish National Science Center Grant Sheng 2018/30/Q/ST2/00811.
\end{acknowledgments}
\bibliography{bibl.bib}

\begin{thebibliography}{13}%
\makeatletter
\providecommand \@ifxundefined [1]{%
 \@ifx{#1\undefined}
}%
\providecommand \@ifnum [1]{%
 \ifnum #1\expandafter \@firstoftwo
 \else \expandafter \@secondoftwo
 \fi
}%
\providecommand \@ifx [1]{%
 \ifx #1\expandafter \@firstoftwo
 \else \expandafter \@secondoftwo
 \fi
}%
\providecommand \natexlab [1]{#1}%
\providecommand \enquote  [1]{``#1''}%
\providecommand \bibnamefont  [1]{#1}%
\providecommand \bibfnamefont [1]{#1}%
\providecommand \citenamefont [1]{#1}%
\providecommand \href@noop [0]{\@secondoftwo}%
\providecommand \href [0]{\begingroup \@sanitize@url \@href}%
\providecommand \@href[1]{\@@startlink{#1}\@@href}%
\providecommand \@@href[1]{\endgroup#1\@@endlink}%
\providecommand \@sanitize@url [0]{\catcode `\\12\catcode `\$12\catcode
  `\&12\catcode `\#12\catcode `\^12\catcode `\_12\catcode `\%12\relax}%
\providecommand \@@startlink[1]{}%
\providecommand \@@endlink[0]{}%
\providecommand \url  [0]{\begingroup\@sanitize@url \@url }%
\providecommand \@url [1]{\endgroup\@href {#1}{\urlprefix }}%
\providecommand \urlprefix  [0]{URL }%
\providecommand \Eprint [0]{\href }%
\providecommand \doibase [0]{http://dx.doi.org/}%
\providecommand \selectlanguage [0]{\@gobble}%
\providecommand \bibinfo  [0]{\@secondoftwo}%
\providecommand \bibfield  [0]{\@secondoftwo}%
\providecommand \translation [1]{[#1]}%
\providecommand \BibitemOpen [0]{}%
\providecommand \bibitemStop [0]{}%
\providecommand \bibitemNoStop [0]{.\EOS\space}%
\providecommand \EOS [0]{\spacefactor3000\relax}%
\providecommand \BibitemShut  [1]{\csname bibitem#1\endcsname}%
\let\auto@bib@innerbib\@empty
\bibitem [{\citenamefont {Kamiński}(2019)}]{Kaminski:2019tqo}%
  \BibitemOpen
  \bibfield  {author} {\bibinfo {author} {\bibfnamefont {W.}~\bibnamefont
  {Kamiński}},\ }\href@noop {} {\  (\bibinfo {year} {2019})},\ \Eprint
  {http://arxiv.org/abs/1906.07554} {arXiv:1906.07554 [gr-qc]} \BibitemShut
  {NoStop}%
\bibitem [{\citenamefont {Bonga}\ and\ \citenamefont
  {Gupt}(2016{\natexlab{a}})}]{Bonga:2015kaa}%
  \BibitemOpen
  \bibfield  {author} {\bibinfo {author} {\bibfnamefont {B.}~\bibnamefont
  {Bonga}}\ and\ \bibinfo {author} {\bibfnamefont {B.}~\bibnamefont {Gupt}},\
  }\href {\doibase 10.1007/s10714-016-2071-0} {\bibfield  {journal} {\bibinfo
  {journal} {Gen. Rel. Grav.}\ }\textbf {\bibinfo {volume} {48}},\ \bibinfo
  {pages} {71} (\bibinfo {year} {2016}{\natexlab{a}})},\ \Eprint
  {http://arxiv.org/abs/1510.00680} {arXiv:1510.00680 [gr-qc]} \BibitemShut
  {NoStop}%
\bibitem [{\citenamefont {Bonga}\ and\ \citenamefont
  {Gupt}(2016{\natexlab{b}})}]{Bonga:2015xna}%
  \BibitemOpen
  \bibfield  {author} {\bibinfo {author} {\bibfnamefont {B.}~\bibnamefont
  {Bonga}}\ and\ \bibinfo {author} {\bibfnamefont {B.}~\bibnamefont {Gupt}},\
  }\href {\doibase 10.1103/PhysRevD.93.063513} {\bibfield  {journal} {\bibinfo
  {journal} {Phys. Rev.}\ }\textbf {\bibinfo {volume} {D93}},\ \bibinfo {pages}
  {063513} (\bibinfo {year} {2016}{\natexlab{b}})},\ \Eprint
  {http://arxiv.org/abs/1510.04896} {arXiv:1510.04896 [gr-qc]} \BibitemShut
  {NoStop}%
\bibitem [{\citenamefont {Agullo}\ and\ \citenamefont
  {Morris}(2015)}]{Agullo:2015tca}%
  \BibitemOpen
  \bibfield  {author} {\bibinfo {author} {\bibfnamefont {I.}~\bibnamefont
  {Agullo}}\ and\ \bibinfo {author} {\bibfnamefont {N.~A.}\ \bibnamefont
  {Morris}},\ }\href {\doibase 10.1103/PhysRevD.92.124040} {\bibfield
  {journal} {\bibinfo  {journal} {Phys. Rev.}\ }\textbf {\bibinfo {volume}
  {D92}},\ \bibinfo {pages} {124040} (\bibinfo {year} {2015})},\ \Eprint
  {http://arxiv.org/abs/1509.05693} {arXiv:1509.05693 [gr-qc]} \BibitemShut
  {NoStop}%
\bibitem [{\citenamefont {Akrami}\ \emph {et~al.}(2018)\citenamefont {Akrami}
  \emph {et~al.}}]{Akrami:2018odb}%
  \BibitemOpen
  \bibfield  {author} {\bibinfo {author} {\bibfnamefont {Y.}~\bibnamefont
  {Akrami}} \emph {et~al.} (\bibinfo {collaboration} {Planck}),\ }\href@noop {}
  {\  (\bibinfo {year} {2018})},\ \Eprint {http://arxiv.org/abs/1807.06211}
  {arXiv:1807.06211 [astro-ph.CO]} \BibitemShut {NoStop}%
\bibitem [{\citenamefont {Ashtekar}\ \emph {et~al.}(2009)\citenamefont
  {Ashtekar}, \citenamefont {Kaminski},\ and\ \citenamefont
  {Lewandowski}}]{Ashtekar:2009mb}%
  \BibitemOpen
  \bibfield  {author} {\bibinfo {author} {\bibfnamefont {A.}~\bibnamefont
  {Ashtekar}}, \bibinfo {author} {\bibfnamefont {W.}~\bibnamefont {Kaminski}},
  \ and\ \bibinfo {author} {\bibfnamefont {J.}~\bibnamefont {Lewandowski}},\
  }\href {\doibase 10.1103/PhysRevD.79.064030} {\bibfield  {journal} {\bibinfo
  {journal} {Phys. Rev.}\ }\textbf {\bibinfo {volume} {D79}},\ \bibinfo {pages}
  {064030} (\bibinfo {year} {2009})},\ \Eprint {http://arxiv.org/abs/0901.0933}
  {arXiv:0901.0933 [gr-qc]} \BibitemShut {NoStop}%
\bibitem [{\citenamefont {Agullo}\ \emph {et~al.}(2013)\citenamefont {Agullo},
  \citenamefont {Ashtekar},\ and\ \citenamefont {Nelson}}]{Agullo:2013ai}%
  \BibitemOpen
  \bibfield  {author} {\bibinfo {author} {\bibfnamefont {I.}~\bibnamefont
  {Agullo}}, \bibinfo {author} {\bibfnamefont {A.}~\bibnamefont {Ashtekar}}, \
  and\ \bibinfo {author} {\bibfnamefont {W.}~\bibnamefont {Nelson}},\ }\href
  {\doibase 10.1088/0264-9381/30/8/085014} {\bibfield  {journal} {\bibinfo
  {journal} {Class. Quant. Grav.}\ }\textbf {\bibinfo {volume} {30}},\ \bibinfo
  {pages} {085014} (\bibinfo {year} {2013})},\ \Eprint
  {http://arxiv.org/abs/1302.0254} {arXiv:1302.0254 [gr-qc]} \BibitemShut
  {NoStop}%
\bibitem [{\citenamefont {Agullo}\ \emph {et~al.}(2012)\citenamefont {Agullo},
  \citenamefont {Ashtekar},\ and\ \citenamefont {Nelson}}]{Agullo:2012sh}%
  \BibitemOpen
  \bibfield  {author} {\bibinfo {author} {\bibfnamefont {I.}~\bibnamefont
  {Agullo}}, \bibinfo {author} {\bibfnamefont {A.}~\bibnamefont {Ashtekar}}, \
  and\ \bibinfo {author} {\bibfnamefont {W.}~\bibnamefont {Nelson}},\ }\href
  {\doibase 10.1103/PhysRevLett.109.251301} {\bibfield  {journal} {\bibinfo
  {journal} {Phys. Rev. Lett.}\ }\textbf {\bibinfo {volume} {109}},\ \bibinfo
  {pages} {251301} (\bibinfo {year} {2012})},\ \Eprint
  {http://arxiv.org/abs/1209.1609} {arXiv:1209.1609 [gr-qc]} \BibitemShut
  {NoStop}%
\bibitem [{\citenamefont {Agullo}\ and\ \citenamefont
  {Singh}(2017)}]{Agullo:2016tjh}%
  \BibitemOpen
  \bibfield  {author} {\bibinfo {author} {\bibfnamefont {I.}~\bibnamefont
  {Agullo}}\ and\ \bibinfo {author} {\bibfnamefont {P.}~\bibnamefont {Singh}},\
  }in\ \href {\doibase 10.1142/9789813220003_0007} {\emph {\bibinfo {booktitle}
  {Loop Quantum Gravity: The First 30 Years}}},\ \bibinfo {editor} {edited by\
  \bibinfo {editor} {\bibfnamefont {A.}~\bibnamefont {Ashtekar}}\ and\ \bibinfo
  {editor} {\bibfnamefont {J.}~\bibnamefont {Pullin}}}\ (\bibinfo  {publisher}
  {WSP},\ \bibinfo {year} {2017})\ pp.\ \bibinfo {pages} {183--240},\ \Eprint
  {http://arxiv.org/abs/1612.01236} {arXiv:1612.01236 [gr-qc]} \BibitemShut
  {NoStop}%
\bibitem [{\citenamefont {Ashtekar}\ \emph {et~al.}(2006)\citenamefont
  {Ashtekar}, \citenamefont {Pawlowski},\ and\ \citenamefont
  {Singh}}]{Ashtekar:2006wn}%
  \BibitemOpen
  \bibfield  {author} {\bibinfo {author} {\bibfnamefont {A.}~\bibnamefont
  {Ashtekar}}, \bibinfo {author} {\bibfnamefont {T.}~\bibnamefont {Pawlowski}},
  \ and\ \bibinfo {author} {\bibfnamefont {P.}~\bibnamefont {Singh}},\ }\href
  {\doibase 10.1103/PhysRevD.74.084003} {\bibfield  {journal} {\bibinfo
  {journal} {Phys. Rev.}\ }\textbf {\bibinfo {volume} {D74}},\ \bibinfo {pages}
  {084003} (\bibinfo {year} {2006})},\ \Eprint
  {http://arxiv.org/abs/gr-qc/0607039} {arXiv:gr-qc/0607039 [gr-qc]}
  \BibitemShut {NoStop}%
\bibitem [{\citenamefont {Martin-Benito}\ \emph {et~al.}(2009)\citenamefont
  {Martin-Benito}, \citenamefont {Marugan},\ and\ \citenamefont
  {Olmedo}}]{MartinBenito:2009aj}%
  \BibitemOpen
  \bibfield  {author} {\bibinfo {author} {\bibfnamefont {M.}~\bibnamefont
  {Martin-Benito}}, \bibinfo {author} {\bibfnamefont {G.~A.~M.}\ \bibnamefont
  {Marugan}}, \ and\ \bibinfo {author} {\bibfnamefont {J.}~\bibnamefont
  {Olmedo}},\ }\href {\doibase 10.1103/PhysRevD.80.104015} {\bibfield
  {journal} {\bibinfo  {journal} {Phys. Rev.}\ }\textbf {\bibinfo {volume}
  {D80}},\ \bibinfo {pages} {104015} (\bibinfo {year} {2009})},\ \Eprint
  {http://arxiv.org/abs/0909.2829} {arXiv:0909.2829 [gr-qc]} \BibitemShut
  {NoStop}%
\bibitem [{\citenamefont {Ashtekar}\ \emph {et~al.}(2008)\citenamefont
  {Ashtekar}, \citenamefont {Corichi},\ and\ \citenamefont
  {Singh}}]{Ashtekar:2007em}%
  \BibitemOpen
  \bibfield  {author} {\bibinfo {author} {\bibfnamefont {A.}~\bibnamefont
  {Ashtekar}}, \bibinfo {author} {\bibfnamefont {A.}~\bibnamefont {Corichi}}, \
  and\ \bibinfo {author} {\bibfnamefont {P.}~\bibnamefont {Singh}},\ }\href
  {\doibase 10.1103/PhysRevD.77.024046} {\bibfield  {journal} {\bibinfo
  {journal} {Phys. Rev.}\ }\textbf {\bibinfo {volume} {D77}},\ \bibinfo {pages}
  {024046} (\bibinfo {year} {2008})},\ \Eprint {http://arxiv.org/abs/0710.3565}
  {arXiv:0710.3565 [gr-qc]} \BibitemShut {NoStop}%
\bibitem [{\citenamefont {Mena~Marugan}\ \emph {et~al.}(2011)\citenamefont
  {Mena~Marugan}, \citenamefont {Olmedo},\ and\ \citenamefont
  {Pawlowski}}]{MenaMarugan:2011me}%
  \BibitemOpen
  \bibfield  {author} {\bibinfo {author} {\bibfnamefont {G.~A.}\ \bibnamefont
  {Mena~Marugan}}, \bibinfo {author} {\bibfnamefont {J.}~\bibnamefont
  {Olmedo}}, \ and\ \bibinfo {author} {\bibfnamefont {T.}~\bibnamefont
  {Pawlowski}},\ }\href {\doibase 10.1103/PhysRevD.84.064012} {\bibfield
  {journal} {\bibinfo  {journal} {Phys. Rev.}\ }\textbf {\bibinfo {volume}
  {D84}},\ \bibinfo {pages} {064012} (\bibinfo {year} {2011})},\ \Eprint
  {http://arxiv.org/abs/1108.0829} {arXiv:1108.0829 [gr-qc]} \BibitemShut
  {NoStop}%
\end{thebibliography}%
\end{document}